\documentclass[3p]{elsarticle}
\newcommand{\beq}{\begin{equation}}
\newcommand{\eeq}{\end{equation}}
\newcommand{\bea}{\begin{eqnarray}}
\newcommand{\eea}{\end{eqnarray}}
\newcommand{\uvec}[1]{{\bf \hat{#1}}}
\newcommand{\eq}[1]{{Eq. (\ref{#1})}}

\newcommand{\ket}[1]{{|{#1}\rangle}}
\usepackage{graphicx}
\usepackage{bm}
\usepackage{amsmath}
\usepackage{amsfonts}
\usepackage{lineno}
\usepackage{hyperref}
\journal{Chaos, Solitons \& Fractals}

\usepackage{numcompress}
\begin{document}

\begin{frontmatter}
\title{Rogue breather modes: Topological sectors, and the `belt-trick', in a one-dimensional ferromagnetic spin chain}

\author[]{Rahul O. R.}
\author[]{S. Murugesh\corref{cor1} }
\ead{murugesh@iist.ac.in}
\address{Department of Physics, Indian Institute of Space Science and Technology, Thiruvananthapuram - 695~547, India.}
\cortext[cor1]{Corresponding author}

\begin{abstract}
We present explicit solutions for breather soliton modes of excitation in the  one-dimensional Heisenberg ferromagnetic spin chain. We identify a characteristic geometrical feature of these breather modes wherein a helicoidal configuration of spins  is continuously transformed to one which differs from the initial helicoid by a  total twist of `2'. This is a curious manoeuvre  popularly known as the `belt trick', an illustration of the simple connectedness of the $\rm{SU}(2)$ group manifold, and its rotation period $4\pi$. We show that this effectively splits the configuration space of the ferromagnetic chain in one-dimension into two topological sectors, distinguished by their total twist --- either `0', or `1'. Further, the energy lower bound of the two sectors is separated by a finite gap varying inversely with the size of the lattice.

\end{abstract}


\end{frontmatter}

\section{Introduction}
The Heisenberg model of ferromagnets (HF), driven by exchange interactions, has been a subject of keen interest for decades. Although fundamentally quantum in nature, for large values of spin, such as the macrospin associated with domains,  its classical counterpart has nevertheless yielded important physically realizable results. The bare Heisenberg model in one-dimension (1-d) is perhaps the simplest yet non-trivial nonlinear model among the larger class of $O(3)$ sigma models\cite{rr:1987,zakr:1989}. In 1-d, the model is completely integrable with soliton solutions and is gauge equivalent to the nonlinear Schr\"{o}dinger equation (NLSE) \cite{ml:1977,takt:1977,zs:1972,zs:1979}. The classical model in 2-d, though not known to be integrable, has a very interesting class of solutions, {\it skyrmions}, whose energies fall into distinct topological sectors identified by the Hopf number\cite{bp:1975}. Stable localized solutions are, however, not possible in the bare Heisenberg model in three and higher dimensions\cite{rr:1987}. Nevertheless, topologically non-trivial spin 
configurations are still possible in higher dimension, aided by constraints, or if the model goes beyond  exchange interaction\cite{niemi:1997,fad:2002,kiv:2012,coop:1999}.

Besides its fundamentality, interest in the bare Heisenberg ferromagnet, especially in 1-d, has seen a phenomenal revival owing to definite technological prospects, particularly {\it magnonics}\cite{hille:2015}. Spin waves, or {\it magnons}, manipulated by either a spin polarized current or current induced magnetic field, have demonstrated a clear possibility of driving next generation memory and computing hardware. With its inherent non-volatility, robustness, speed and minimal energy consumption, magnonics offers a clear alternative in realizing our persistent thirst for faster, smaller, yet more energy efficient memory devices.   

In the study of integrable systems and solitons the classical HF, in 1-d especially, is an important model owing to its complete integrability\cite{ml:1977,takt:1977}. Although in reality, the dynamics of a spin chain is likely to be driven by a volley of other factors, such as applied fields, spin polarized currents, inherent anisotropy, and shape effects in the form of a demagnetizing field, the solutions to the base model remain the primary driver for understanding the extended system. The HF admits soliton solutions --- localized and  robust spin excitations of finite energy, traveling at uniform speed. In this paper we discuss another class of spin waves, {\it breathers}, which are also solitons in that they are an outcome of the integrability of the HF model, and obtained via the inverse scattering transformation method. However, as we show, the breathers are phenomenally different from solitons, and also display a curious topological side in the process of their evolution. Particularly, we show that the breather mode is essentially the curious manoeuvre variedly known as { `Dirac's string trick', the `belt trick'} or the { `Balinese plate trick'}, wherein a spin chain with ends fixed is continuously transformed to one which differs from the initial configuration by a  total twist `2'.  This is frequently considered an illustration of the simple connectedness of the $\rm{SU(2)}$ group manifold,
--- a double cover on an open ball of radius $\pi$ with diametrically opposite points identified,  
and its period $4\pi$. In understanding the physics of the electron spin, the $4\pi$ periodicity of the $\rm{SU(2)}$  group is especially of immense significance connecting the spin magnetic moment of the electron to its {\it spinor} representation. However, the topological property illustrated in the following sections is that of a  {\it classical} spin chain, wherein continuity and  periodic boundary conditions are the primary players responsible.  In effect, this demonstrates a division of the configuration space of the continuous spin chain into two topologically distinct sectors, determined by the total twist --- either 0 or 1, respectively.
It should be emphasized herein that while the HF model refers to a spin chain in 1-d, the spin field itself lives in an internal space that is three dimensional, which is essential for exhibiting such a non-trivial geometry. To our knowledge, this is perhaps the first 
real world illustration of the `belt-trick' where it  arises naturally, in this case as an exact solution for the Heisenberg ferromagnetic spin chain. In the context of magnonics, such a spin breather is a potential  candidate for a memory element --- the total twist, for instance, indicating its logical state, which can be possibly written or erased by a breather mode excitation. 

The subject of breather modes in a ferromagnetic chain has had a fair share of attention in literature, although recent. For instance, breathers
	associated with magnonic seed solutions to the HF have been obtained in \cite{zai:2016}.  Exploiting the gauge equivalence of the HF to the NLSE, in \cite{Mukh:2015} Mukhopadhyay, et. al., investigate its behavior by tracking its energy density. Higher order extensions of the ferromagnetic chain, in relation to higher order NLSE, have also been investigated using the same technique\cite{sun:2015}. Through a detailed study using inverse scattering transform technique, Demontis et. al.,  \cite{demo:2018} have recently shown that breather like behavior can be imitated by a bound state of two or more soliton modes, which 
	individually would asymptotically approach a uniform magnetization otherwise.  Breather modes in discrete versions of HF have also been investigated in the past\cite{khal:2003}. However, the 
 results presented below assume significance in that we reveal an intrinsic topological connection to the breather spin mode, hitherto unrealized. It should be emphasized here that the model being studied here is one dimensional, wherein such a non-trivial topology  is usually unexpected. Moreover, we present explicit analytical expressions for the spin breather modes, which have not been presented before.
\section{Breather modes in the 1-D HF}
\label{sec:breathermode}
The classical 1-d Heisenberg ferromagnet (HF) is governed by the hamiltonian
\beq\label{hf}
\mathbf{H} = J\int\bigg{(}{\frac{\partial\uvec{S}}{\partial x}}\bigg{)}^2\,dx,
\eeq
where $J$ is a constant that depends on the exchange integral, positive for a ferromagnet,  $\uvec{S}(x)=\{S_1,S_2,S_3\}$ is the normalized unit spin field ($|\uvec{S}|=1$) of interest, and $x$ parameterizes the 1-d lattice. Applying the variational principle, or using spin commutation relations and then applying the continuum limit, one gets the simplest form of the Landau-Lifshitz equation (LLE) governing its dynamics (after rescaling $t$)
\beq\label{ll}
\uvec{S}_t  = \uvec{S}\times\uvec{S}_{xx},
\eeq
where the subscripts stand for partial derivatives, as usual. 
 
The 1-d LLE, \eq{ll}, is integrable through inverse scattering method with soliton solutions. The most prominent solution of the Heisenberg model is a one-soliton of the `secant-hyperbolic' form in a 1-d infinite chain\cite{ml:1977,takt:1977,fad:1987}. This is a localized traveling spin wave of finite energy, vanishing asymptotically. Although the spin field $\uvec{S}$ is presented as a vector field in \eq{ll}, its true  description is more complete only when represented as spinors, and inverse scattering transformation methods bring out this character quite naturally. However, while the `secant-hyperbolic' 1-soliton solution is indeed non-trivial, it fails to bring out the inherent features of a spinor. We show below that the {\it breather}, is characteristically distinct, and especially that it brings out the true spinor character of the spin field.

As the spinor association is central to comprehending the intricate geometry of the spin chain dynamics, we give a brief outline on how
it arises naturally in inverse scattering methods for an integrable system. As mentioned earlier, the 1-d LLE, \eq{ll}, is integrable through the method of inverse
scattering transforms, possess soliton solutions, and can be expressed
through the auxiliary linear equations\cite{fad:1987}
\beq
\label{lap}
\begin{aligned}
\bm{\Phi}_x &= U_{HF} \bm{\Phi} = i\lambda{\bf S} \bm{\Phi} \\
\bm{\Phi}_t &= V_{HF} \bm{\Phi} = \Big{(}\lambda{\bf S}{\bf S}_x +2i\lambda^2{\bf S}\Big{)} \bm{\Phi}.
\end{aligned}
\eeq
Here
\beq\label{spin}
{\bf S} = \sum_i S_i\sigma_i,
\eeq
with $\sigma_i,\,\,i=1,2,3,$ being the Pauli matrices, $S_i$ the components of the unit spin field $\uvec{S}$, \eq{hf}, and $\lambda$  the scattering parameter. $U_{HF}$ and $V_{HF}$ form the Lax pair for the system.  The compatibility of the two equations, in \eq{lap}, $\bm{\Phi}_{xt}=\bm{\Phi}_{tx}$, gives the matrix form of the LLE:
\beq
{\bf S}_t = \frac{-i}{2}[{\bf S}, {\bf S}_{xx}],
\eeq
which, when expressed in component form, gives back \eq{ll}. 
The spin field ${\bf S}$, as expressed
in \eq{spin}, and the Lax pair in \eq{lap} are  elements of $\rm{su}(2)$ lie algebra (taking $\lambda$ to be real). 
Consequently, the auxiliary function $\bm{\Phi}$ is an element of the
$\rm {SU}(2)$ group. 
The spin field ${\bf S}$ itself can be written in the form 
\beq\label{map}
{\bf S} = \lim_{\lambda\to 0} \bm{\Psi}^{\dagger}\sigma_3 \bm{\Psi},
\eeq
for an appropriate choice of $\bm{\Psi} \in \rm{SU}(2)$\cite{schief:2002}. The method of inverse scattering transforms provides a systematic way of finding the unitary matrix $\bm{\Psi}$ for each soliton solution (See Appendix A for details). Being an element of $\rm{SU}(2)$, $\bm{\Psi}(x,t)$ represents
a rotation matrix in $\mathbb{C}^2$. For a given value of $x$ and $t$ (and $\lambda=0$), each such $\bm{\Psi}$ can be represented by a point in the topological space of $\rm{SU}(2)$ --- a double cover on an open ball of radius $\pi$, with antipodal points identified\cite{choquet:1982}. 
Finding soliton solutions for the LLE is then equivalent to 
finding an appropriate $\bm{\Psi}$ (\eq{map}). Given a {\it seed} solution ${\bf S}$,
a further solution ${\bf S'}$ can be found using established methods, such as a Darboux transformation\cite{schief:2002}, denoted $\mathbf{P}$, with
\beq
\bm{\Psi}' = \mathbf{P}\bm{\Psi} \in \rm{SU}(2),
\eeq
and ${\bf S}'= \lim_{\lambda\to 0}\bm{\Psi}^{'\dagger}\sigma_3\bm{\Psi}'$ (See Appendix A for details). 
While the LLE, \eq{ll}, may describe a general spatially distributed spin mode and its time evolution,  soliton modes obtained through inverse scattering transforms are special --- their distribution and time evolution are described by appropriate matrix functions that live in $\rm{SU}(2)$, and hence describe a spinor evolution. 
As we shall show below, it is this spinor association, and its nontrivial topology,  that is revealed by the breather spin mode, one that the usual soliton fails to do. 
Although much of our further discussion will be in the language of the spin field $\uvec{S}$, the association with the corresponding $\rm{SU}(2)$ element is omnipresent and exact. 


Breather modes are obtained if one starts with a  seed solution of the form 
\beq\label{s0}
\uvec{S}_0=\cos(2\kappa_0x)\uvec{y} + \sin(2\kappa_0x)\uvec{z}
\eeq
where $\kappa_0$ is an arbitrary real constant, and $4\kappa_0^2$ the energy density.  This is perhaps the simplest spatially inhomogeneous spin mode possible for the LLE, and an 
elementary case of a general larger class suggested by
Lakshmanan, et. al. in \cite{ml:1976}.  Two more constants, arising out of the global rotation and translation symmetry of the Heisenberg model, have been taken to be zero without loss of generality. This solution corresponds to a spatially periodic, static, spin field whose energy scales linearly with the size of the system. The locus of the tips of adjacent  spin vectors is a helicoid with axis along the lattice direction $\uvec{x}$. In order to keep the energy finite, we shall consider a lattice of size $L$, with periodic boundary conditions. Thus $0< x\le L$, and $\kappa_0 = n\pi/L$, where $n$ is any integer.   With this identification, the spin chain may be thought of as continuously  embedded on a circular lattice, and whose time evolution is a smooth deformation of the spin mode. The state space of the spin chain is then $S^2\times S^1$. The seed, \eq{s0}, then corresponds to a total energy $E_0 = (2n\pi)^2/L$. 
 
Starting from \eq{s0} and proceeding with, say, the Darboux method\cite{schief:2002}, one obtains, after some tedious yet straight forward algebra, the breather mode (See Appendix A for details):

\bea\label{s1}
\uvec{S}_1(x,t) = \frac{\lambda_{0R}^2-\lambda_{0I}^2}{|\lambda_0|^2} \uvec{S}_0
+ \Bigg{[} \frac{2\, \lambda_{0I}^2}{|\lambda_0|^2\chi^2} \xi\eta
- \frac{2\lambda_{0I}\lambda_{0R}}{|\lambda_0|^2\chi	} \zeta 
\Bigg{]}\uvec{x} \, \nonumber \\
+ \Bigg{[} \frac{2\, \lambda_{0I}^2}{|\lambda_0|^2 \chi^2} \, \xi \, 
		\Big( \xi \, \cos(2\kappa_0 x) - \zeta \sin(2\kappa_0 x)  \Big)
- \frac{2\, \lambda_{0I} \lambda_{0R}}{|\lambda_0|^2 \chi} \, \eta \, \sin(2\kappa_0 x) \, 
		\Bigg{]}\uvec{y} \, \nonumber \\
+ \Bigg{[} \frac{2\, \lambda_{0I}^2}{|\lambda_0|^2 \chi^2} \, \xi \, 
		\Big( \zeta \, \cos(2\kappa_0 x) + \xi \sin(2\kappa_0 x)  \Big)
- \frac{2\, \lambda_{0I} \lambda_{0R}}{|\lambda_0|^2 \chi} \, \eta \, \cos(2\kappa_0 x)\, 
		\Bigg{]}\uvec{z} \, 
\eea
where,\\
 $\zeta = c_1\cos(\Omega_{0R}) +c_2\cosh(\Omega_{0I}),
 $\\
 $\eta = c_3\sin(\Omega_{0R}) -c_4\sinh(\Omega_{0I}),
 $\\
 $\xi = c_4\sin(\Omega_{0R}) +c_3\sinh(\Omega_{0I}),
 $\\
 $\chi = c_2\cos(\Omega_{0R}) +c_1\cosh(\Omega_{0I}),
 $\\
 $\Omega_0 = \Omega_{0R}+ i\Omega_{0I} = 2 p_0(x+2\lambda_0t)$,\\
 $p_0 = p_{0R} + ip_{0I} = \sqrt{\kappa_0^2 + {\lambda_0}^2} $, \\
 $c_1=2(\lambda_{0I}^2+p_{0R}^2-\lambda_{0R}p_{0R}-\lambda_{0I}p_{0I})$,\\
 $c_2=2\kappa_0(p_{0I}-\lambda_{0I})$,\\
 $c_3=-2(\lambda_{0R}^2+p_{0I}^2-\lambda_{0R}p_{0R}-\lambda_{0I}p_{0I})$,\\
 $c_4=2\kappa_0(p_{0R}-\lambda_{0R})$.\\
Here, it may be verified that $\eta^2+\zeta^2+\xi^2 = \chi^2$, and $c_2^2+c_3^2+c_4^2 = c_1^2$. 
 The form and behavior of the breather mode is dictated by two parameters --- $\kappa_0$ introduced in \eq{s0}, and $\lambda_0 (=\lambda_{0R}+i\lambda_{0I})$ the complex eigenvalue on the plane of the scattering parameter, in the language of inverse scattering transform theory. Generally, this spin mode is localized. 
The relative values of the two parameters, $\kappa_0$ and $\lambda_0$, lead to broadly three distinct behavior --- periodic in time, space, or both, validating its `breather' nomenclature. However, as is clear from the form of the solution, \eq{s1}, it is cumbersome to analyze the  behavior in its entire generality. For our further analysis, therefore, we shall consider only the special case wherein  the eigenvalue is purely imaginary, $\lambda_0=ia$, where $a$ is positive (as $\lambda_0$ can assume values only on the upper half of the complex plane of the scattering parameter\cite{spn:1984}). It suffices to note that this restriction nevertheless captures all the qualitative features of the breather. With a purely imaginary $\lambda_0$, the spin field in \eq{s1} takes a simpler form
\bea\label{s1r}
\uvec{S}_1(x,t)= -\uvec{S}_0
 	+\frac{2\xi}{\chi^2}\bigg{[}%
 	 \eta \, \uvec{x}
 	+\Big( \xi\cos(2\kappa_0x)-\zeta\sin(2\kappa_0x) \Big)
 	\uvec{y}
 	+\Big( \zeta\cos(2\kappa_0x) + \xi\sin(2\kappa_0x) \Big)
 	 \uvec{z} \bigg{]}
\eea
 	Here, $\uvec{S}_0$ is the seed solution given in \eq{s0}, and $\xi$, $\chi$, $\eta$ and $\zeta$ were defined below \eq{s1}.

Broadly, there are two distinct cases i) $\kappa_0<a$ and  ii) $\kappa_0>a$. 
The more interesting situation is indeed the case when $\kappa_0\sim a$ which leads to a certain {\it rogue} behavior, which we shall address in detail a little later:

\noindent {\bf Case i --- $\kappa_0<a$:}
Firstly, it can be  directly verified that as $\kappa_0\to 0$, the form of the breather soliton, \eq{s1}, approaches the `secant-hyperbolic' isolated one-soliton. This is easier understood from the seed solution, \eq{s0}, which reduces to a constant field for $\kappa_0=0$. For $\kappa_0<a$, from  the expressions below \eq{s1}, $p_0=\sqrt{\kappa_0^2-a^2}$ is purely imaginary. Thus $\Omega_{0R}$ is a pure function of time, while
$\Omega_{0I}$ is a function only of $x$. From the expression for $\xi$, $\chi$, $\eta$ and $\zeta$ we are lead to conclude that  the spin field is localized in space, and periodic in time, with a period 
\beq\label{period1}
T=\frac{\pi}{2a\sqrt{a^2-\kappa_0^2}}.
\eeq
However, it does not satisfy our condition of  spatial periodicity, and hence is disallowed.  

\noindent {\bf Case ii --- $\kappa_0>a$:}
The variables defined below \eq{s1} reduce to the following:
\bea\label{vari}
 \zeta = 2\kappa_0^2\cos(\Omega_{0R}) -2\kappa_0a\cosh(\Omega_{0I}),
\nonumber\\
\eta = -2\kappa_0\sqrt{\kappa_0^2-a^2}\sinh(\Omega_{0I}),
\nonumber\\
\xi = 2\kappa_0\sqrt{\kappa^2_0-a^2}\sin(\Omega_{0R}),\nonumber \\
\chi = -2\kappa_0a\cos(\Omega_{0R}) +2\kappa_0^2\cosh(\Omega_{0I}),
\nonumber\\
\Omega_0 = \Omega_{0R}+ i\Omega_{0I} = 2 \sqrt{\kappa_0^2-a^2}(x+2iat).
\eea
At $\kappa_0=a$  the spin field is reduced to the background seed field that we started with. However, as  $a$ makes the cross over from $a>\kappa_0$ to $a<\kappa_0$, we see an interesting change in the dynamics of the spin configuration. The temporal periodicity witnessed earlier is replaced by a spatial periodicity, while more interestingly, the field is now temporally local (as a consequence of $p_0$ turning real, in contrast to Case i above). Spatially, the field is composed of two periodic functions with generally non-commensurate periods
\beq\label{period2}
L_1 = \frac{\pi}{\kappa_0},\,\, {\rm and}\,\, L_2 = \frac{\pi}{\sqrt{\kappa_0^2-a^2}}=\gamma L_1,
\eeq
where $\gamma=(1-a^2/\kappa_0^2)^{-1/2}$. The breather vanishes for both $a=\kappa_0$, as pointed out above,  and also for $a=0$, leaving us with the strict case $0<a<\kappa_0$. 
Imposing periodicity of the spin field, $\uvec{S}_1(x+L,t)=\uvec{S}_1(x,t)$, along with the earlier condition $\kappa_0=n\pi/L$, implies $L_2=L/m$ for some positive integer $m$. Thus, 
from \eq{period2},  $n=\gamma m$, or equivalently 
$a= \frac{\pi}{L}\sqrt{n^2-m^2}$.
Neither of these two field configurations (for case i and ii) possess a  traveling nature owing to the fact that we have chosen $\lambda_0$ to be purely imaginary for ease in analysis. A more  general complex $\lambda_0$, while retaining the overall qualitative properties of the breather, renders the filed a uniform speed determined by $\lambda_{0R}$,  as is expected of a soliton.  

The LLE, \eq{ll}, is also gauge equivalent to the integrable non-linear Schr\"{o}dinger equation (NLSE) in {1-d}\cite{ml:1977,takt:1977, zs:1972,zs:1979}. Thus, corresponding to every solution of the LLE there exists one to the NLSE. In particular, in the language of NLSE the breathers for $\kappa_0<a$ (case i) are the time periodic Kuznetsov-Ma breathers\cite{kuz:1977,ma:1979}, while those for $\kappa_0>a$ are the spatially periodic Akhmediev\cite{akhm:1986} breathers. An intermediate limiting case, when both these time and spatial periods ($T$ and $L_2$, in Eqs. (\ref{period1}) and (\ref{period2})) tend to infinity, often referred to as {\it rogue} waves\cite{pere:1983}, are indeed special. The rogue wave is marked by a sudden, momentary, yet colossal enhancement in the field magnitude, localized in space. It has been a subject of keen investigation since observation of similar phenomena in deep ocean waves\cite{sulem:1999, kharif:2008}, wherein the behavior can indeed be described approximately by the NLSE. Possibilities of  rogue phenomena in the atomic scale have also been theoretically suggested in Bose-Einstein condensates\cite{akhm:2009}. For reasons that will be clear immediately, of particular interest in our study is the spin breather obtained by setting  $m=1$, giving the simplest breather with least possible energy over and above that of the seed. It also corresponds to the largest value possible for $\gamma$, given $n$. 
From equations (\ref{period1}) and (\ref{period2}), we identify that a rogue corresponds to the case when $|\kappa_0^2-a^2|\sim 0$. However, in our finite system of size $L$,  the closest one can get to this condition is when $\gamma$ is largest,  i.e., when $m=1$ for any given $n$. Consequently, we shall identify this as the {\it rogue spin mode} in the finite ferromagnetic spin chain. 
\begin{figure}[ht]
\centering
\includegraphics[width=0.6\textwidth]{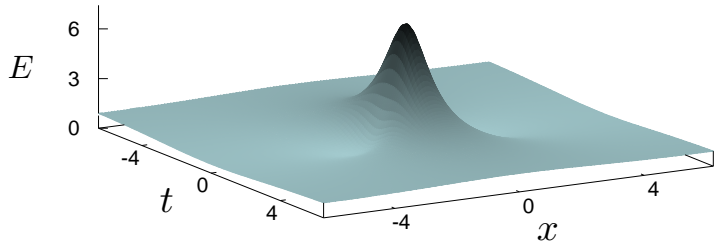}
\caption{The energy density ${E}(x,t)=(\uvec{S}_x)^2$ in the $x-t$ plane for $m=1$ and $n=2$ (or $\kappa_0=2\pi/L$ and $a=\sqrt{3}\pi/L$) ---  rogue spin mode, \eq{s1r}, marked by a momentary colossal excitation local in both space and time. } 
\label{colossal}
\end{figure}
A further justification of this nomenclature is evident from the space time profile of the energy density  in Fig. \ref{colossal}. As is characteristic of a rogue, the energy density clearly shows a sudden colossal rise,  localized in both space and time.
We emphasize that the breather presented here is not a bound state
of multiple regular solitons, unlike in \cite{demo:2018}. They are indeed
one-soliton solutions with a periodic boundary condition, associated with a single complex eigenvalue in the language of inverse scattering transforms.
\section{The rogue breather and the `belt trick'}
\begin{figure}[ht]
\centering
\includegraphics[trim=0cm 0cm 0cm 0.5cm,width=1\textwidth]{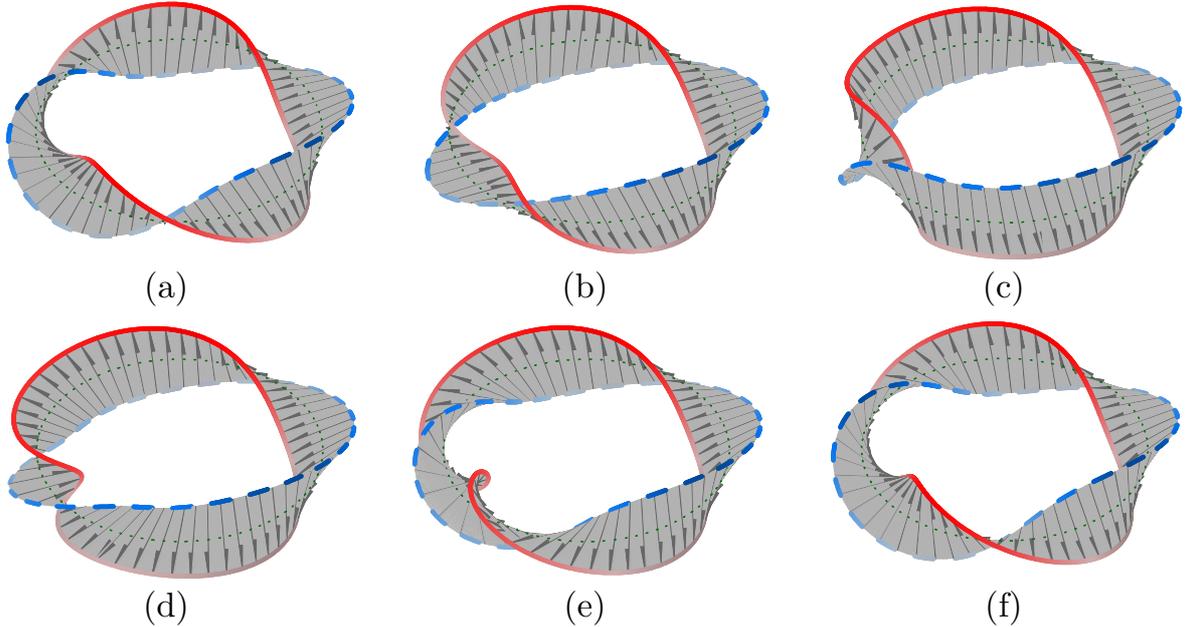}
\caption{Snapshots of the spin configuration	($n=2$ and $m=1$) at different instances through its evolution.  Due to spatial periodicity, the spins at the boundary are identified, making the lattice  a circle. The loci of the top(bottom) ends of the spin vectors are indicated by the red solid lines(blue dashed lines). In (a), (b), (e) and (f), the blue and red loops are {\it linked}, with a linking number `2', while in the intermediate phase, (c) and (d), they are {\it un-linked}. Or, equivalently, the spin field goes through a net twist of `2' (or $4\pi$ radians)  in figures (a), (b), (e) and (f). The total twist in (c) and  (d), however, is zero.   }
\label{spin_conf}
\end{figure}

The characteristic distinction of the breather from a regular soliton mode is evident if we take a closer look at the rogue spin mode as it evolves in time. Recalling
$L_1=L/n$ and $L_2=L/m$ (\eq{period2}), where $m$ and $n$ are positive integers, and the condition that $\kappa_0>a$, we see that
the lowest possible value for $m$ is `1' and that of $n$ is `2'. Simply put, breathers are not possible if we have chose a seed with $\kappa_0=\pi/L$. We discuss in detail the rogue breather mode obtained by setting $m=1$ and $n=2$ (equivalently, $\kappa_0=2\pi/L$ and $a=\sqrt{3}\pi/L$). For purpose of illustration we shall represent the spin vectors as arrows whose centers are located on the corresponding lattice site. Further, the lattice itself can be closed into a circle, owing to the periodic boundary condition (See Fig. \ref{spin_conf}). 
Two closed loops can be imagined --- one formed by the locus of the upper tips of the arrows, and another by the bottom tips. Thus, the time evolution of the spin field is  equivalently described by the twisting and writhing of a closed ribbon bounded by these two loops about the fixed circular lattice\cite{crick:1976}. In Fig. \ref{spin_conf}((a)-(f)) we show a stroboscopic plot of the spin configurations at six sequential instances from its time evolution (see supplementary material \cite{belt} for detailed animation). A curious aspect of this time evolution is that if one considers the spin configurations in Figs. \ref{spin_conf}(a) and \ref{spin_conf}(c), they differ by a total twist of `2'. In its evolution through one period, the twist on the ribbon completes a  cycle, going from `2' to `0', and then back to `2'. The evolution of the spin array is continuous, i.e., the angle between adjacent spin vectors changes continuously, nevertheless the total twist  
exhibits a singular shift. 
 
\begin{figure}[ht]
\centering
\includegraphics[width=0.6\textwidth]{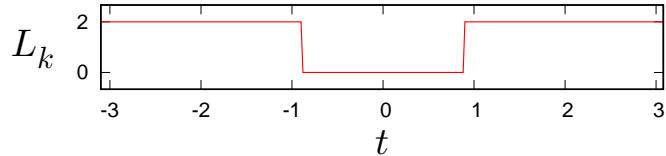}
\caption{A plot of the linking number between the two loops in Fig. \ref{spin_conf}, as a function of time. The behavior of the total twist of the spin field about the lattice direction is identical.}
\label{linking}
\end{figure}

Alternately, this behavior can also be visualized in terms of the linking number, $L_k$, between the two loops described in the previous paragraph, formed by the loci of the top and bottom tips of the spin vectors (see Fig. \ref{linking}). As is evident from Fig. \ref{spin_conf}, the two loops are linked in (a), (b), (e) and (f). In the intermediate phase, (c) and (d), however, they are not. We identify the intermediate phase (Fig. \ref{spin_conf}(c) and \ref{spin_conf}(d)) with the rogue phase in Fig. \ref{colossal}. This result is consistent with the Calugareanu theorem\cite{calu:1959}, later discovered independently by White\cite{white:1969} and Fuller\cite{fuller:1971}:
\beq\label{ltw}
\mathbf{Link = Twist + Writhe}.
\eeq
By construction, the two loops that form the boundary of  the ribbon in Fig. \ref{spin_conf} reduce to the plane circular lattice in the limit $|\uvec{S}|\to 0$, which has zero writhe. Link and twist change simultaneously by $\pm2$ in the course of evolution, while the writhe remains zero throughout\cite{fuller:1978}, satisfying \eq{ltw}.

The rogue breather then is a continuous transformation connecting two spin chain configurations differing by a total twist of 2 (or $4\pi$ radians). This manoeuvre is popularly referred to as the `belt trick' (`Dirac's string trick', `plate trick', etc.,) and is frequently used as an illustration of the $4\pi$ periodicity of the complex rotation group $\rm{SU}(2)$, the group of rotations that define the two dimensional complex spinor space\cite{choquet:1982}. 
It is also a signature to the fact that the topological space of the group of rotations in 3-d, $\rm{SO}(3)$, is not simply connected. We explore further by tracking the indicatrix --- the curve traced by the vector $\uvec{S}(x)$ at any instant of time on the surface of the unit sphere where it lives. Due to spatial periodicity, the indicatrix forms a smooth closed loop on the surface of this unit sphere. The area of this loop, or flux due to the indicatrix, on the surface  of the sphere then is given by 
\beq\label{area}
\mathbf{F}_I= \int{\uvec{S}\cdot {\bf da}} = \oint{{\bf A}\cdot {\bf dl}}.
\eeq 
Here, we have invoked Stokes' theorem for the closed loop traced by $\uvec{S}$ on the surface of the unit sphere, and  ${\bf A}$ is the vector potential, such that
$\nabla\times{\bf A}=\uvec{S}$. The vector potential ${\bf A}$ may then be likened to that of a magnetic monopole of magnetic charge $4\pi/\mu_0$. 
Using spherical polar co-ordinates to write $\uvec{S} = (\sin\theta\cos\phi,\sin\theta\sin\phi,\cos\theta)$, the vector potential is  given by\cite{saku:1994}
\beq\label{vp}
{\bf A} = \frac{(\pm 1 - \cos\theta)}{\sin\theta}\uvec{\phi},
\eeq
with the `$\pm$' applicable in upper and lower hemispheres, respectively, both nevertheless leading to the same area.  
Substituting in \eq{area}, we get 
\beq\label{area2}
\mathbf{F}_I = \oint \frac{\partial\phi}{\partial x}{(1 - \cos\theta)}dx.
\eeq
The LLE, \eq{ll}, can be rewritten for the conjugate variables $\phi$ and $U=\cos\theta$ as
\beq
\begin{aligned}
\dot{\phi} &= -\frac{U_{xx}}{(1-U^2)} - \frac{U_x^2U}{(1-U^2)^2} - \phi^2_{x}U \\
\dot{U} &= \big(\phi_x(1-U^2)\big)_x,
\end{aligned}
\eeq
using which it can be shown, after some straight forward algebra, that the area $\mathbf{F}_I$ is  a constant. Alternately, one may use another interesting connection, namely that the area $\mathbf{F}_I$ in \eq{area2}, covered by the closed indicatrix loop, is in fact the  total momentum of the spin chain, and is the generator for spatial translations\cite{tjon:1977}.  Since a uniform lateral shift of the spin mode does not change its total energy, the total momentum commutes with the Hamiltonian in \eq{hf} and hence remains a constant. With $m=1$ and $n=2$ for the rogue spin mode, using \eq{s1r} a direct calculation of the integral in \eq{area2} yields $\mathbf{F}_I=4\pi$. 
In Fig. \ref{indicatrix} we illustrate the time evolution of the indicatrix curve for the rogue breather mode (see supplementary material\cite{indic} for detailed animation). The curve intersects itself once. With the intersection preserved, the area enclosed by the indicatrix on the surface of the unit sphere remains constant.   
\begin{figure}[b]
\centering
\includegraphics[trim = 0 0 0 0,width=0.9\textwidth]{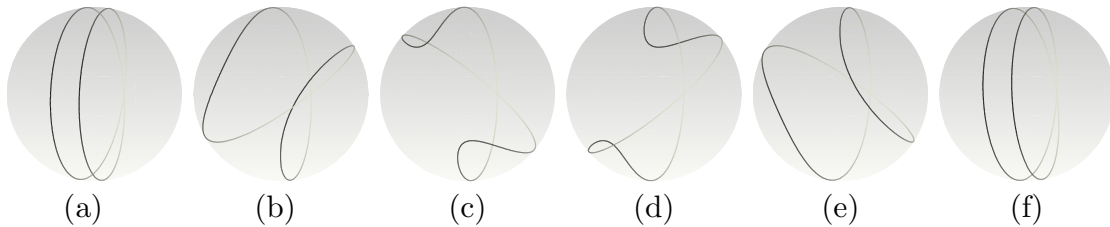}
\caption{Snapshots of the closed loop on the unit sphere due to the indicatrix traced by the spin vector for the breather spin mode ($m=1$, $n=2$). The indicatrix intersects itself once, which is preserved through the time evolution. The curve is symmetric about the great circle passing through the intersection.  Bearing in mind the direction of the loop, the total area covered by the loop on the surface of the sphere remains a constant ($4\pi$).}
\label{indicatrix}
\end{figure}

For larger values of $n$, the total twist of the seed solution increases (infact, ${\bf Twist}=n$ for the seed mode). A breather with $m=1$ then will be a rogue that would alter the total twist from $n\to n-2$, and back. However, in this paper we shall desist from a more detailed discussion on breather spin modes with higher values of $n$ and $m$, and shall confine ourselves only to an exposition of the simplest rogue mode and its intricacies.
\section{Spinor association}
To recognize the spinor nature of the breather mode it is essential to realize that, while the indicatrix describes the configuration of the spin chain alone, there is additional information provided by the $\rm{SU(2)}$ element one obtains in the
process of inverse scattering transform (see \eq{map}). As $\rm {SU(2)}$ is homomorphic to $\rm {SO(3)}$ --- the group of
rotations in 3-d, one may imagine the evolution as 
describing an orthogonal triad of vectors including $\uvec{S}$. The Euler angles describing the triad  form the three parameters for the $\rm {SO(3)}$ matrix element thus obtained, or equivalently it may thought of as defining an axis of rotation and an angle of rotation. 
The parameter space of these three variables is an open ball of radius $\pi$, with antipodal points {\it identified}, and 
any spin chain with periodic boundary conditions forms a closed loop within the sphere. In \eq{map}, we showed the relation connecting the spin field $\uvec{S}$ with the $\rm {SU(2)}$ element that naturally arises in the inverse scattering theory. If one were to write such a 
matrix as $\bm{\Psi}=e^{-i(\uvec{n}\cdot\vec{\bm\sigma})\Theta/2}$, then this matrix can be represented in the parametric sphere by a point at a radial distance $\Theta$ in the direction $\uvec{n}$\cite{choquet:1982}. 

\begin{figure}[h]
\centering
  \includegraphics[trim = 0 0 0 0,width=1\textwidth]{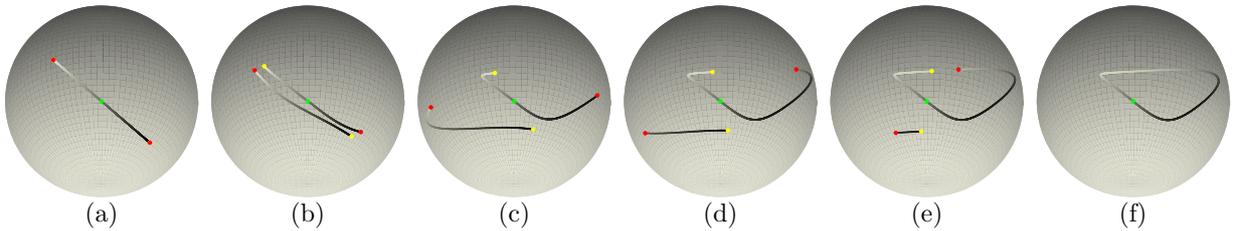}
	\caption{The breather mode as seen in the group manifold --- an open sphere of radius $\pi$, with antipodal points identified.
		 For $m=1$ and $n=2$ (twist `2'), the seed solution, \eq{s0}, corresponds to two overlapping diametrical lines. The antipodal pairs are indicated by dots of the same color. In the process of the breather mode evolution the two lines join to form a closed loop within the sphere (f), which can be then shrunk to a point continuously (See supplementary material \cite{topo} for detailed animation). The same cannot be achieved for a twist `1' loop --- a single line connecting two antipodal points. }
	\label{SO3}
\end{figure}

For the seed solution \eq{s0}, it can be shown that (See Appendix A for details)
\begin{equation}\label{psi0}
\lim_{\lambda\to 0}\bm{\Psi}_0(x,t,\lambda) =  	\begin{pmatrix}
\cos(\kappa_0 x-\frac{\pi}{4})\, e^{i(\kappa_0^2t-\frac{\pi}{4})} & i\sin(\kappa_0 x-\frac{\pi}{4})\, e^{i(\kappa_0^2t-\frac{\pi}{4})} \\i\sin(\kappa_0 x-\frac{\pi}{4})\, e^{-i(\kappa_0^2t-\frac{\pi}{4})} \, & \, \, \cos(\kappa_0 x-\frac{\pi}{4})\, e^{-i(\kappa_0^2t-\frac{\pi}{4})}
\end{pmatrix} = e^{i(\uvec{z}\cdot\vec{\bm{\sigma}})(\kappa_0^2t-\frac{\pi}{4})}e^{i(\uvec{x}\cdot\vec{\bm{\sigma}})\Theta/2},
\end{equation}
with $\Theta/2+\frac{\pi}{4} = \kappa_ox=\frac{n\pi x}{L}$. 
Applying in \eq{map}, we find the spin field in matrix form
\beq
{\bf S}_0(x,t)=e^{-i(\uvec{x}\cdot\vec{\bm{\sigma}})\Theta/2}e^{-i(\uvec{z}\cdot\vec{\bm{\sigma}})(\kappa_0^2t-\frac{\pi}{4})}\sigma_3e^{i(\uvec{z}\cdot\vec{\bm{\sigma}})(\kappa_0^2t-\frac{\pi}{4})}e^{i(\uvec{x}\cdot\vec{\bm{\sigma}})\Theta/2} =e^{-i(\uvec{x}\cdot\vec{\bm{\sigma}})\Theta/2}\sigma_3e^{i(\uvec{x}\cdot\vec{\bm{\sigma}})\Theta/2} = {\bf S}_0(x,0),
\eeq
the seed solution, as given in \eq{s0}.
In the parametric sphere, at $t=\pi/4\kappa_0^2$ this seed solution, \eq{psi0}, for $n=1$ ($0\le x<L$) is then a line running from end to end diametrically, along the $\uvec{x}$ direction, signifying a  total twist of `1' (or $2\pi$ radians).
From the form of ${\bm \Psi}_0$ in the last term in \eq{psi0}, we infer that in a time period $2\pi/\kappa_0^2(=2L^2/\pi)$, this diametric line completes one full global rotation.  For $n=2$, $0\le \Theta<4\pi$, corresponds  to  two overlapping diametrical lines --- a total twist  `2'. For the breather mode in \eq{s1}, we have
\beq\label{psi1}
\lim_{\lambda\to 0}{\bm \Psi}_1(x,t,\lambda) = \lim_{\lambda\to 0} \Big( \frac{1}{\sqrt{d_1}}\ \mathbf{P}_1{\bm \Psi}_0 \Big), 
\eeq 
where $\mathbf{P}_1(x,t,\lambda)$ is the Darboux matrix with (See Appendix A for details)
\beq \label{P1}
\lim_{\lambda\to 0} \Big( \frac{1}{\sqrt{d_1}} \mathbf{P}_1 \Big) =\frac{1}{\chi}
\begin{pmatrix}
	-i\xi & -e^{2 i \kappa_0^2 t} (\eta +i\zeta) \\
	e^{-2 i \kappa_0^2 t} (\eta -i\zeta ) & i\xi
\end{pmatrix},
\eeq
with $\chi$, $\xi$, $\eta$ and $\zeta$ as in \eq{vari}, 
and ${\bm\Psi}_0$ given in \eq{psi0}. Since $\xi^2+\eta^2+\xi^2=\chi^2$, we may define variables $\alpha$ and $\beta$ such that
\beq
\sin\alpha\cos\beta = \frac{\eta}{\chi},\,\,
\sin\alpha\sin\beta = \frac{\zeta}{\chi},\,\,\,
\cos\alpha = \frac{\xi}{\chi},
\eeq
so that \eq{P1} may be rewritten
\beq\label{p12}
\lim_{\lambda\to 0} \Big( \frac{1}{\sqrt{d_1}} \mathbf{P}_1 \Big) =
\begin{pmatrix}
	-i\cos\alpha & -\sin\alpha \, e^{i (\beta+2\kappa_0^2 t)} \\
	\sin\alpha \, e^{-i(\beta+2 \kappa_0^2 t)} & i\cos\alpha 
\end{pmatrix},
\eeq
Visualized as a rotation, right hand side of \eq{p12} is a rotation about the axis $\uvec{n}(x,t) = \{\sin\alpha \sin(\beta+2\kappa_0^2t), \sin\alpha \cos(\beta+2\kappa_0^2t),\cos\alpha\}$, but peculiarly through a constant angle $\pi$.  However, the direction $\uvec{n}(x,t)$ is obscured by the complicated form of the variables in \eq{vari}.   
Fig. \ref{SO3} illustrates the time evolution of the rogue breather, \eq{psi1}, as seen in this parameter space. 
In the process of the breather evolution, the two diametrical lines (as in the seed solution, for $n=2$) evolve into a closed loop within the sphere (Fig. \ref{SO3} (f)). This closed loop is topologically equivalent to a point to which it can be continuously shrunk. However, the HF being a conservative system, such a shrinkage is prevented by the constraint on energy. 
In the process of evolution the loop then retraces its path back to the two overlapping diametrical lines as in Fig. \ref{SO3} (a) (See supplementary material \cite{topo} for detailed animation). Such a closed loop can indeed be continuously shrunk to a point, for instance, when dissipation effects are incorporated in the model. 
While a loop corresponding to a total twist `2' can be continuously shrunk to one with twist `0', the same cannot be achieved for a loop with twist `1', as the antipodal pair of points cannot be delinked in any continuous transformation --- this  is indeed the gist of the belt trick. Consequently, we may conclude that the collection of all periodic breather modes form two topologically distinct classes distinguished by the nature of their total twist --- `odd', or `even'.  

\section{Energy bounds and topological sectors}
Lastly, we import the consequence of a rather well known result from the theory of curves --- Fenchel's theorem\cite{fench:1929,kamien:2002}, for the breather modes. First, we note that the energy of the breather is bounded from below for any non-zero $\kappa_0$ (see \eq{s0}), owing to the condition of periodicity\cite{dand:2015}. The HF is invariant under global rotations, which is reflected in the constancy of total spin 
\beq
{\bf S}_{tot} = \oint{\uvec{S}\,dx}.
\eeq
Particularly, the spatially periodic breathers we have discussed are a set of modes corresponding to the sector ${\bf S}_{tot}=0$. 
Given the unit norm 
of the spin field vector, one may associate integral curves, ${\bf R}(x)$, whose tangent vector is given by $\uvec{S}(x)$, such that 
\beq
\frac{d{\bf R}}{dx}\equiv \uvec{S}.
\eeq
I.e., we interpret ${\bf R}(x)$ as a non-stretching curve in three dimension, such that `$x$' is its arc length parameter\cite{struik:1961}. The curvature of such a curve is given by $|\partial\uvec{S}/\partial x|$, and   the condition ${\bf S}_{tot}=0$ then implies that the curve ${\bf R}(x)$ is closed. Besides, ${\bf S}_{tot}$ being `0', the indicatrix (see Fig. \ref{indicatrix}, Section 3) traced by $\uvec{S}$ on the unit sphere cannot be confined to a single hemisphere, any which way one looks at it. Fenchel's theorem then follows from this statement\cite{kamien:2002}: 
\beq\label{fenc1}
\oint|\frac{\partial\uvec{S}}{\partial x}|dx \ge 2\pi.
\eeq    
Notice that $|\frac{\partial\uvec{S}}{\partial x}|^2$ is the energy density of the spin chain. Invoking Cauchy-Schwarz inequality on \eq{fenc1} one can arrive at a lower limit on the total energy of the spin configuration\cite{dand:2015}:
\beq
E_{Total} \ge 4\pi^2/L. 
\eeq 

 The HF model we have investigated is free of any dissipation. If dissipation effects, such as Gilbert damping\cite{gilb:2004} are to be included, 
our presentation illustrates that any mode with total twist $T_w$ will be continuously transformed to one with a total twist of $T_w-2$, and eventually to a twist of either `0' or `1',  mediated by breather  excitations. Thus we obtain a more general lower bound for the total energy of the spin chain with periodic boundary condition, determined by its twist:
\begin{equation}
E_{Total} \ge \left\{
\begin{array}{c l}
0 & \text{ for\,\, `even'\,\, twist},\\
4 \pi^2/L & \text{ for\,\, `odd' \,\, twist}.
\end{array} \right. 
\end{equation}
 Consequently, the configuration space of the spin chain,  with  a given value for $\uvec{S}_{tot}$, is divided into two topologically distinct sectors differing by a total twist of `1' ---  also an exemplification of the fact that the fundamental group of $\rm{SO}(3)$ is $\mathbb{Z}_2$. It is indeed interesting that
 an exact solution for the HF model exits, notably, a soliton solution in the form of {\it breather} modes as a mediator shifting the total twist of the chain by `2'.
\section{Conclusion}
In conclusion, we have identified a breather spin mode for the Heisenberg ferromagnet, and have  revealed an exotic side to the geometry of the spin chain excitation seldom captured by a regular soliton.  
Specifically, the breather mode, a temporally local spin excitation,  is a practical illustration of the curious manoeuvre well known as the `belt-trick', or the `Dirac string trick', often used as a demonstration of the simple connectedness of the $\rm{SU}(2)$ group manifold. In the context of the breather spin mode, the belt-trick is a continuous transformation of the spin chain wherein the total twist is changed by `2'.  We have also pointed out that the total energy of the spin chain has a lower bound, depending inversely on the size of the chain. If dissipation effects were to be regarded, the configuration space is broadly reduced to two topological sectors with distinct energy lower bounds, determined by their total twist --- `odd' or `even'. 
\section*{Acknowledgments}
We thank Dr. Naveen Surendran for helpful discussions. This research did not receive any specific grant from funding agencies in the public, commercial, or not-for-profit sectors.
\appendix
\section{Breather spin modes using Darboux transformation technique}
In order to find an explicit analytic expression for the  breather by Darboux transform technique, one may take as a starting point the seed solution \eq{s0} and  the Lax pair for HF, \eq{lap}. Instead, we utilize the gauge equivalence of NLSE to the HF to arrive at the same result. I.e., we start with an equivalent seed solution for the NLSE. Using the Lax pair for NLSE, we first obtain the breather soliton through a Darboux transformation.
Then, using its gauge equivalence to HF, we obtain the breather spin mode.  This procedure has the advantage that it also provides us with the matrix function $\bm{\Psi}$ defined in \eq{map}, which is essential for Section 4.
We give a brief outline of the procedure and the Darboux method below.  
 
The integrable NLSE for the complex field $\psi(x,t)$,
\begin{equation}
\label{app:nlse}
	i \psi_t + \psi_{xx} + 2 |\psi|^2 \psi = 0 ,
\end{equation}
arises as the compatibility condition $\bm{\Psi}_{xt}=\bm{\Psi}_{tx}$ for the system of linear differential equations,
\begin{equation}
\label{app:lpnlse}
\begin{aligned}
\bm{\Psi}_x &= U \bm{\Psi}, \quad \bm{\Psi}_t &= V \bm{\Psi}.
\end{aligned}
\end{equation}
The connections $U(x,t,\lambda)$ and $V(x,t,\lambda)$, its Lax pair, are given by
\begin{equation}
U = \begin{pmatrix}
							 0 & \psi \\ -\overline{\psi} & 0 
			 \end{pmatrix}
			 + \lambda			 
			 \begin{pmatrix}
							 -i & 0 \\ 0 & i
			 \end{pmatrix} 			, \quad
V = \begin{pmatrix}
							 i{|\psi|}^2 & i \psi_x \\ i \overline{\psi}_x & -i{|\psi|}^2 
			 \end{pmatrix}
			 + \lambda			 
			 \begin{pmatrix}
							  0 & 2\psi \\ -2\overline{\psi} & 0
			 \end{pmatrix} 
			 + \lambda^2			 
			 \begin{pmatrix}
							 -2i & 0 \\ 0 & 2i
			 \end{pmatrix}  
\end{equation}
and $\lambda$ being the scattering parameter.  This system is gauge equivalent to the LLE\cite{zs:1979}, \eq{ll}, with the Lax pair given in \eq{lap}. $\bm{\Psi}$ is a matrix valued auxiliary function, usually imagined as representing a fictitious wave scattered by a {`potential'} $U$.  
For a given $\bm{\Psi}$, the corresponding spin field ${\bf S}$ is then found using the relation in \eq{map}:
\beq\label{map2}
{\bf S} = \lim_{\lambda\to 0} \bm{\Psi}^\dagger\sigma_3\bm{\Psi}.
\eeq
For the temporally periodic solution to  NLSE,
\begin{equation}
\label{app:seed}
\psi_0(x,t) = \kappa_0\, e^{2 i \kappa_0^2 t},
\end{equation}
with a real constant $\kappa_0$, the auxiliary matrix function  $\bm{\Psi}_0(x,t,\lambda)$, can be found through direct integration as,
\begin{equation}
\label{app:psi0}
\bm{\Psi}_0(x,t,\lambda) =   \frac{1}{\sqrt{d_0}}\	\begin{pmatrix}
						\varphi_1 & -\overline{\varphi}_2 \\ \varphi_2 \, & \, \, \overline{\varphi}_1
						\end{pmatrix} 
\end{equation}
where,
\begin{equation}\label{varphi}
	\begin{aligned}
	\varphi_1 	&= \Big( e^{-i \Omega/2} + i\, \frac{(\lambda - p)}{\kappa_0}\, e^{i \Omega/2}  \Big) 
														\, e^{i \kappa_0^2 t},  	\\
	\varphi_2  	&= \Big( i\, \frac{(\lambda - p)}{\kappa_0}\, e^{-i \Omega/2} + e^{i \Omega/2}  \Big) 
														\, e^{-i \kappa_0^2 t}, 	\\
	\Omega 		&= 2\, p ( x + 2 \lambda t ) , \quad
	p 	  		= \sqrt{\kappa_0^2 + \lambda^2 } , \quad
	d_0 		= 4 ( \kappa_0^2 + \lambda^2 - \lambda p) /  \kappa_0^2.
	\end{aligned}
\end{equation}
Substituting \eq{app:psi0} in \eq{map2}, the corresponding spin field can be seen to be
\beq\label{app:s0}
{\bf S} = \cos(2\kappa_0 x)\sigma_2 + \sin(2\kappa_0x)\sigma_3, 
\eeq
the seed solution we chose in \eq{s0}. 

The auxiliary matrix function for the breather, $\bm{\Psi}_1$, is then 
related to  $\bm{\Psi}_0$ through a Darboux transformation $\mathbf{P}_1$:
\beq
\bm{\Psi}_1= \frac{1}{\sqrt{d_1}}\ \mathbf{P}_1\bm{\Psi}_0. 
\eeq
where $d_1 = \text{det}\, \mathbf{P}_1 $.
If $\ket{\psi}_i$, $i=1,2$ are two column vectors such that
\beq
\begin{aligned}
\ket{\psi}_{ix} &= U(\lambda_i)\ket{\psi}_i \\
\ket{\psi}_{it} &= V(\lambda_i)\ket{\psi}_i,
\end{aligned}
\eeq
for two complex eigenvalues $\lambda_1$ and $\lambda_2$, then 
the Darboux transformation $\mathbf{P}_1$ is defined such that
\beq\label{pdef}
\mathbf{P}_1(\lambda_i)\ket{\psi}_i=0
\eeq
and 
can be written as\cite{schief:2002}
\beq\label{pq}
\mathbf{P}_1 = \lambda\mathbf{I} +\mathbf{Q}_1.
\eeq
For the NLSE, $\lambda_1= \overline{\lambda}_2=\lambda_0$, and
\beq\label{kets}
\ket{\psi}_1 = \begin{pmatrix}
	\varphi_1(\lambda_0)\\
	\varphi_2(\lambda_0)
\end{pmatrix},\hskip .5cm
\ket{\psi}_2=\begin{pmatrix}
	-\overline{\varphi}_2(\lambda_0)\\
	\overline{\varphi}_1(\lambda_0)
\end{pmatrix},
\eeq
where $\varphi_i$ were defined in \eq{varphi}.
Substituting \eq{kets} in \eq{pdef}, and using the expression for
$\mathbf{P}_1$ in \eq{pq}, we obtain, after some straight forward calculation
\begin{equation}
\label{app:P1}
\mathbf{P}_1 = 	\begin{pmatrix}
				\lambda - \lambda_{0R} & 0 \\
				 0 & \lambda - \lambda_{0R}
				\end{pmatrix} 
		+ i\, \frac{ \lambda_{0I}}{\chi} 
		\begin{pmatrix}
		-\xi & -e^{2 i \kappa_0^2 t} (\zeta - i\, \eta ) \\
	 	-e^{-2 i \kappa_0^2 t} (\zeta + i\, \eta ) & \xi
		\end{pmatrix}
\end{equation}
where, $\kappa_0$ is the same real constant appearing in Eq. \eqref{app:s0}, $\lambda_0=\lambda_{0R}+i\lambda_{0I}$ is an arbitrary complex number. Functions $\zeta, \eta, \xi$ and $ \chi$ were defined below Eq. \eqref{s1}. From Eq. \eqref{app:P1} one can find that,  $ d_1 = \text{det}\, \mathbf{P}_1 = (\lambda^2 + |\lambda_0|^2 - 2\lambda \lambda_{0R})$.

\section*{References}
\bibliography{mybibfile}
\end{document}